\documentclass[runningheads]{llncs}
\usepackage{graphicx}
\usepackage{amsmath,amssymb} %
\usepackage{color}
\usepackage{multirow}

\usepackage{amsthm, amsfonts}

\usepackage{subfigure}
\mathchardef\mhyphen="2D

\DeclareRobustCommand{\eg}{\emph{e.g. }}

\DeclareRobustCommand{\ie}{\emph{i.e. }}

\DeclareRobustCommand{\etal}{{et al. }}

\begin{document}
\title{Prediction of Thrombectomy Functional Outcomes using Multimodal Data}
\titlerunning{Prediction of Thrombectomy Functional Outcomes using Multimodal Data}

\author{Zeynel A. Samak\inst{1} \orcidID{0000-0002-3835-4811}\and
Philip Clatworthy \inst{2,3} \orcidID{0000-0002-1206-3573}\and
Majid Mirmehdi\inst{1} \orcidID{0000-0002-6478-1403}
}

\authorrunning{Samak et al.}
\institute{Department of Computer Science, University of Bristol, Bristol, UK 
\and Translational Health Sciences, University of Bristol, Bristol, UK 
\and
Stroke Neurology, Southmead Hospital, North Bristol NHS Trust, Bristol, UK
\email{\{zeynel.samak, phil.clatworthy, m.mirmehdi\}@bristol.ac.uk} 
}

\maketitle              %
\setcounter{footnote}{0}

\begin{abstract}
Recent randomised clinical trials have shown that patients with ischaemic stroke {due to occlusion of a large intracranial blood vessel} benefit from endovascular thrombectomy. However, predicting outcome of treatment in an individual patient remains a challenge. We propose a novel deep learning approach to directly exploit multimodal data (clinical metadata information, imaging data, and imaging biomarkers extracted from images) to estimate the success of endovascular treatment. We incorporate an attention mechanism in our architecture to model global feature inter-dependencies, both channel-wise and spatially. We perform comparative experiments using unimodal and multimodal data, to predict functional outcome (modified Rankin Scale score, mRS)  and achieve 0.75 AUC for dichotomised mRS scores and 0.35 classification accuracy for  individual mRS scores.
\end{abstract}


\section{Introduction}
\label{sec:intro}

Acute stroke is the second most common cause of death worldwide \cite{WHO2018TheDeath}  and the fourth in the UK \cite{StrokeAssociation2018StateStatistics}. More than 80\% of strokes are ischaemic \cite{renowden2014imaging}, caused by a blockage of a blood vessel in the brain by a blood clot, leading to lack of oxygen and death of brain tissue.  Mechanical thrombectomy, is the most effective treatment for the most severe and potentially disabling ischaemic strokes, caused by occlusion of large cerebral blood vessels. In this treatment, thrombi in the large intracranial arteries are mechanically removed via an intra-arterial catheter {and blood flow is restored}.

Recent randomised clinical trials have shown that patients treated with endovascular thrombectomy (EVT) have a greatly increased chance of being independent 3 months after stroke \cite{GOYAL20161723}. In clinical trials functional independence is generally measured using the modified Rankin Scale score (mRS), ranging from 0 (no symptoms) through increasing levels of symptoms and dependency to 6 (death).

Thrombectomy carries a risk of adverse effects, including brain haemorrhage and a small risk of death. Treatment is therefore carried out according to the treating clinician’s judgement about the estimated risks and benefits for each individual patient. The aim is to deploy all available information (\ie imaging and clinical data) to decide on the most favourable treatment option as soon as possible, as treatment effect is strongly anchored on time to treatment.

{Different clinical trials of thrombectomy have used different imaging methods along with more or less stringent criteria for inclusion, and} selecting the patients most suitable  for  treatment remains a challenge. Estimating the potential success of EVT would support clinical decision-making, improving access to treatment for those who might benefit  and potentially avoiding treatment where risks are highest. 

Over the last decade, there has been a rise in the popularity of deep learning methods due to their promising success in various problems (\ie image classification, object detection and medical image analysis). Convolutional neural networks (CNNs) in particular, have been applied to numerous medical image analysis problems and have achieved state-of-the-art results on tasks such as brain tumour segmentation \cite{kamnitsas2017ensembles} and disease prediction \cite{parisot2018disease}. There have been numerous attempts to apply machine and deep learning techniques in the area of automated ischaemic stroke {\it lesion detection}, {\it segmentation} and {\it prognosis} \cite{kamnitsas2017ensembles,lisowska2017context,HILBERT2019103516}, including  the Ischaemic Stroke Lesion Segmentation (ISLES)\footnote{\url{http://www.isles-challenge.org}} Challenge \cite{maier2017isles}. Most of the studies have focused on either segmentation of stroke lesions on baseline imaging or predicting tissue outcome from `baseline' to `follow-up' scans. Although segmentation and detection of ischaemic stroke could assist the clinical-decision process, it requires  voxel-wise annotation of huge quantities of data which is laborious and costly. Also, segmentation does not provide direct information about the success rate of treatment. 

Different imaging modalities are available for evaluation of stroke in the context of thrombectomy treatment selection \cite{fahed2018dwi,albers2018thrombectomy}. The most widely used and available are non-contrast CT (NCCT) and CT angiography (CTA). The MR CLEAN study \cite{fransen2014mr} was the first randomised controlled trial to demonstrate the effectiveness of thrombectomy for ischaemic stroke and used only NCCT and CTA images. We have used the dataset from the MR CLEAN trial in this study.

In this paper, we propose a novel CNN architecture to process raw baseline NCCT image volumes {(the first scan when the patient was admitted to hospital)} and clinical metadata to predict the functional outcome of EVT with either dichotomised or individual mRS scores {3 months after treatment}. Two attention modules, spatial and channel attention,  based on Squeeze and Excitation \cite{Hu2018Squeeze-and-ExcitationNetworks} are used in our architecture which encourage the network to focus on important regions and for more accurate outcome predictions. We explore pre-trained networks and focal loss to handle class imbalance in our dataset.

{To make more accurate treatment outcome predictions, it is important to train our learning model with patients who have received treatment, as well as those who have not. }
Therefore, note that in the rest of this paper, when we state that a network is trained with 3D NCCT image volume only, clinical metadata only, or fusion of both, each case also includes {\it treatment information}, defined as a simple one-hot vector of whether a patient was randomized for EVT or not.

To the best our knowledge, there is no existing method that can predict outcome of EVT from multimodal data, \ie 3D NCCT volumes and clinical metadata. The main contributions of our work are as follows. We introduce a CNN architecture that is able to handle multimodal data, \ie radiological image and clinical metadata information, to estimate the degree of success if EVT is performed.
We integrate an attention mechanism in our architecture to capture discriminatory features both spatially and channel-wise. Extensive comparative experiments are conducted on different architectures, including a CNN network \cite{Bacchi2019DeepStudy}, the encoder part of 3D U-Net \cite{cciccek20163d} which is a popular network in the medical image analysis community, and 3D ResNet networks with and without
pre-training from MedicalNet \cite{chen2019med3d}, to demonstrate the significance of our proposed network. Our multimodal method obtains the best performance in comparative experiments, with 0.75 AUC and 0.62 {\it F1-score} in dichotomous  mRS scores and 0.35 classification accuracy in individual mRS scores. 

The rest of the paper is structured as follows. We review the related works in Section \ref{sec:related_work} and review the MR CLEAN dataset and our preprocessing methods in Section \ref{sec:data}. In Section \ref{sec:methods}, we develop our proposed network and training methods. We explain our experimental set-up and discuss our comparative results in Section \ref{sec:exp_results}. Finally, we provide our conclusion and future work in Section \ref{sec:conclusions}. 
 
\section{Related Works}
\label{sec:related_work}
Early approaches in automated stroke detection and segmentation have relied on simple thresholding methods with standard statistical, correlation-based data analysis, such as  \cite{matesin2001rule,Chawla2009AImages,Rekik2012MedicalAppraisal}, region growing \cite{Rekik2012MedicalAppraisal,Boers2013AutomatedStroke,gupta2014brain} and classical machine learning methods, such as random forest (RFs) \cite{Maier2016PredictingForests,mckinley2017fully,bohme2018combining} and support vector machines (SVMs) \cite{Maier2014IschemicClassifiers}. More recently, CNN-based approaches have proved increasingly popular in solving the ischaemic stroke segmentation challenge, such as \cite{choi2016ensemble,isensee2017brain,lisowska2017context,kamnitsas2017ensembles,pinto2018enhancing,Winzeck2018ISLESMRI.}. Next, we categorise examples of existing works by the data modality applied to automatic stroke detection and/or prediction.

{\bf Clinical Metadata Analysis --} Many existing works for the prognosis of stroke treatment are based on clinical data only, investigating  machine learning methods, such as logistic regression (LR) \cite{weimar2002predicting,Venemaj1710,heo2018machine}, SVMs \cite{Asadi2014,Bentley2014,vanOs2018PredictingAlgorithms}, RFs \cite{Maier2016PredictingForests,heo2018machine,vanOs2018PredictingAlgorithms}, artificial neural networks (ANNs) \cite{Asadi2014,vanOs2018PredictingAlgorithms} and deep neural networks (DNNs) \cite{heo2018machine}. For example, Weimar \etal \cite{weimar2002predicting} and Kent et al. applied LR on clinical metadata, such as patient demographics, NIHSS scores, diabetes, to predict the outcome of ischaemic stroke patients. Heo \etal \cite{heo2018machine} investigated and compared the applicability of LR, RFs and DNNs on clinical data to predict the outcome of the ischaemic stroke with a large cohort of 2604 patients They found that DNNs performed best (at $0.88$ AUC) on their dataset. Van \etal \cite{vanOs2018PredictingAlgorithms} applied the Super Learner, an ensemble method that finds the optimally weighted combination of predictions of the RF, ANN and SVM algorithms, to predict 3-months functional outcome after endovascular treatment. In Nishi et al. \cite{nishi2019predicting}, LR, RF and SVM were applied to clinical metadata, such as NIHSS score, blood glucose level and  the affected brain side to predict the outcome of ischaemic stroke with large vessel occlusion before applying thrombectomy.

{\bf Image Data Analysis --} There have been relatively few studies for outcome of ischaemic stroke treatment using only imaging data. Maier \etal \cite{Maier2016PredictingForests} first trained a model to predict lesion outcomes from magnetic resonance imaging (MRI) images by extracting statistical (10 percentile values, standard deviation, variance) and shape (region area, perimeter, roundness) features from three regions of interest (ROI) (core lesion, tissue around lesion and the rest of the brain) based on lesion outcome probability maps. Then, they applied RFs to predict ischaemic stroke outcome. 
Hilbert \etal \cite{HILBERT2019103516} employed a ResNet model which was trained with CTA images from the MR CLEAN Registry \cite{jansen2018endovascular} dataset. They determined the functional outcome from a single 2D image, obtained from  a maximum intensity projection of a 3D CTA volume (of the patient). They predicted good or bad outcomes only rather than more fine-grained levels of functional outcomes.  

{\bf Multimodal Data Analysis --} The works in this category, in one way or another, have applied both imaging data of various modalities and clinical metadata for their decision-making process. SVMs were employed by Bentley et al.  \cite{Bentley2014} to distinguish between conditions favourable to thrombolysis treatment or not, based on raw unenhanced CT images and NIHSS scores.
Forkert \etal \cite{forkert2015multiclass} applied 12 SVM models in predicting functional outcome at 30 days after treatment. They used values of lesion overlap from different brain regions and stroke laterality alongside other clinical metadata, such as NIHSS and patient age.
Choi \etal \cite{choi2016ensemble} proposed a CNN model that processed 3D MRI diffusion and perfusion imaging and clinical metadata information, such as time since stroke, time to treatment, and thrombolysis in cerebral infarction, to predict treatment outcome. Additionally, they combined their CNN model with  logistic regression on the clinical metadata alone, as an ensemble model for final prediction. Bacchi \etal  \cite{Bacchi2019DeepStudy} predicted binary outcomes of thrombolysis treatment using a dataset of 204 patients with NCCT image volumes and related clinical data (\eg age, gender, blood pressure, NIHSS scores). Their best performing model (0.74 accuracy, 0.69 {it F1-score}, 0.75 AUC) was a combination of CNN and ANN networks that were trained on both NCCT image volumes and clinical metadata.

\section{Data \& Preprocessing}  
\label{sec:data}
\begin{figure}[!ht]
  \centering
  \subfigure[on hospital admission]{\includegraphics[width=0.30\textwidth]{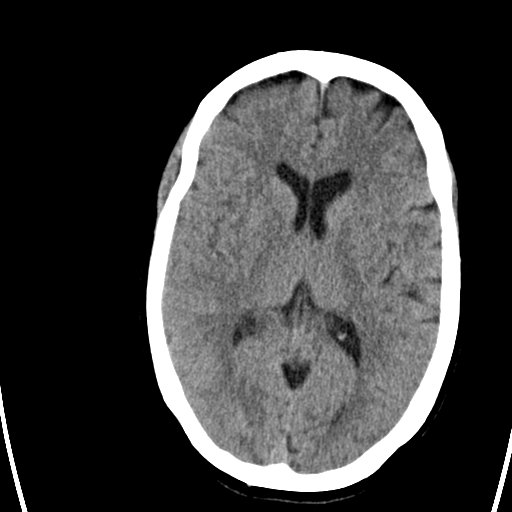}\label{fig:first_scan}}
  \subfigure[24-hr follow-up]{\includegraphics[width=0.30\textwidth]{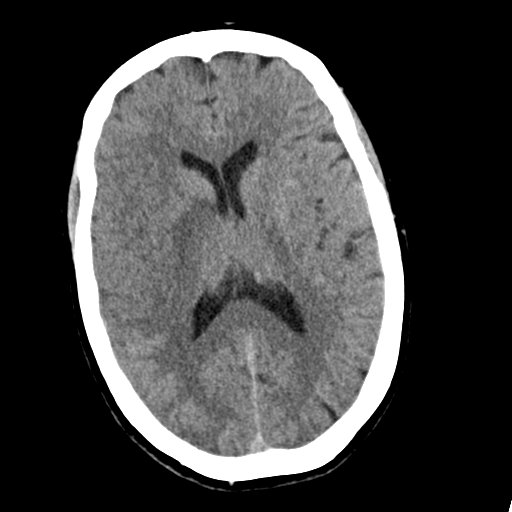}\label{fig:fw24h}}
  \subfigure[1-week follow-up]{\includegraphics[width=0.30\textwidth]{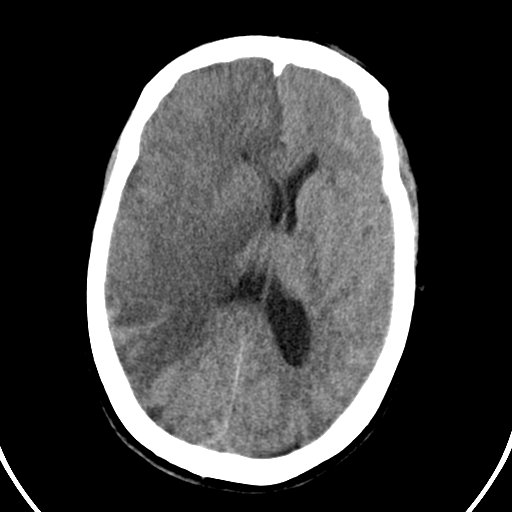}\label{fig:fw1w}}
  \caption{Example Stroke NCCT scans (from MR CLEAN), a) first scan when the patient was admitted to hospital,  b) 24 hour follow-up scan c) 1-week follow-up scan.}
  \label{fig:first_scan-fw24h-fw1w}
\end{figure}

Other than the comprehensive MR CLEAN Trial dataset\footnote{https://www.mrclean-trial.org/home.html} used in this work, there are no other publicly available datasets that contain both image, NCCT image volumes and metadata of patients treated for ischaemic stroke. MR CLEAN was a randomised, clinical trial of intra-arterial treatment and usual care versus usual care in patients with a proximal arterial occlusion in the anterior circulation treated within 6 hours of symptom onset.  Five hundred patients (233 assigned to EVT and 267 to usual care) were treated in 16 medical centres in the Netherlands. The dataset includes a baseline (i.e., the first scan when the patient was admitted to hospital) NCCT and CTA (for all 500 patients), 24-hour follow-up  NCCT and CTA scans (for 394 patients), and additionally a 1-week follow-up NCCT scan (for 358 patients). An example of ischaemic changes on NCCT following a stroke for this timeline (admission, 24-hours, 1 week) can be seen in Fig.\ref{fig:first_scan-fw24h-fw1w}. Additionally, MR CLEAN dataset contains clinical metadata information on its 500 patients, consisting of 90 features, such as patient demographics (age, gender), medical history and stroke scores (NIHSS), imaging biomarkers (ASPECT) and outcome data mRS. For more detailed information about the dataset, we refer the reader to the MR CLEAN study protocol \cite{Berkhemer2015AStroke,fransen2014mr}.

There are various acquisition protocols of the NCCT scans in MR CLEAN due to it being a multi-centre dataset. Through preprocessing, we reduce some of this variation, to allow our learning network to deal with the same standard input and smaller image size. First, all the scans are re-sampled to the same voxel size of 1x1x5$mm^3$ followed by clipping the intensity range between 40 to 100 HU. Then, we crop unnecessary non-brain regions to generate images of size 32x192x192. Also, due to scarce data and to account for variations, data augmentations such as horizontal/vertical flip, rotations, elastic deformations and Gaussian noise are applied (for the network training stage only). The image voxels are finally normalised to zero mean and one standard deviation.

\section{Proposed Method}  \label{sec:methods}

We now present our proposed multimodal network to predict the functional outcome for EVT at multiple levels of the mRS score. We also  explain the methods we used to overcome class imbalance in the dataset, and outline the implementation details of our methods and experiments.

\begin{figure}[t!]
    \centering
    \includegraphics[width=\textwidth]{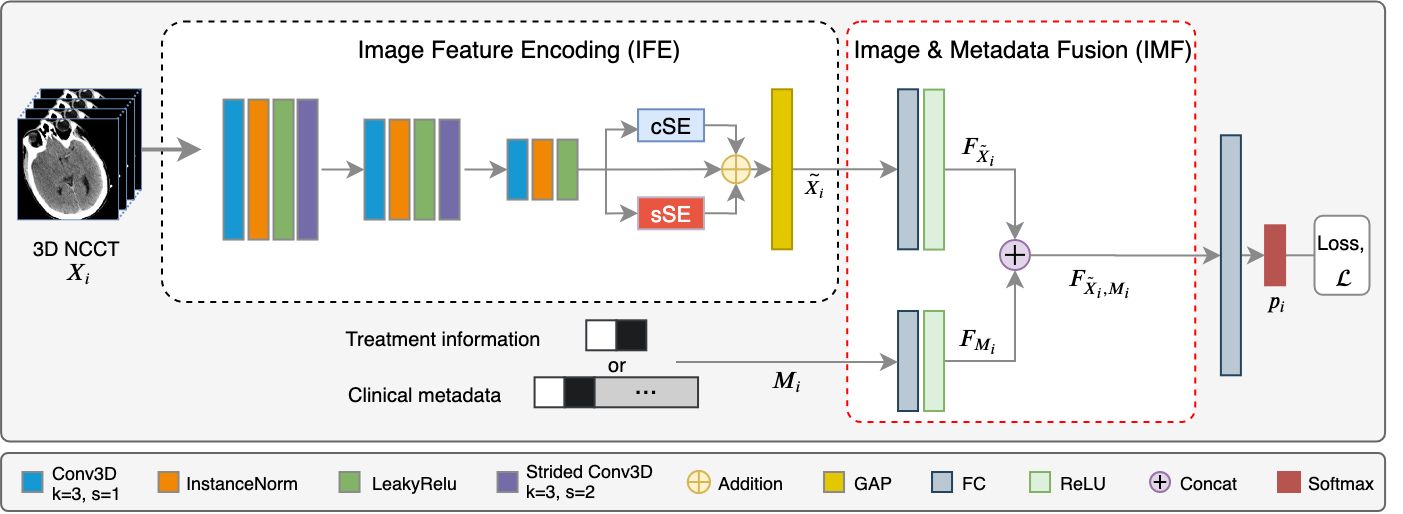}
    \caption{Overview of our model. cSE: channel squeeze and excitation, sCE: spatial squeeze and excitation, \textit{k} is kernel size, \textit{s} is stride size and $\mathcal{L}$ is focal loss.}
    \label{fig:mr-network-design}
\end{figure}

In order to replicate what a clinician would consider, we use multimodal data, \ie NCCT volumes and clinical metadata. In the case of the image volumes, during training, we additionally incorporate a one-hot vector for treatment information only, indicating whether the patient was treated with EVT or not. For 
clinical metadata, we use  patient demographics, medical history, treatment information, radiological image biomarkers (\ie ASPECT score, occlusion, collateral score and symptom side) and medical status/records (hypertension, glucose level etc.) at time of admission to hospital. In the proposed method, we extract NCCT volume features in our {\it image feature encoding} block  (IFE), which later in the experiments section, we compare against the approach in Bacchi \etal \cite{Bacchi2019DeepStudy}, {a U-Net encoder,  and} variations on ResNet. Our image volume features are then combined with metadata information in our {\it image and metadata fusion}  block (IMF) for final mRS prediction. An overview of the proposed network architecture is presented in Fig. \ref{fig:mr-network-design}.

Our outcome prediction of treatment is formulated as follows. Let $Q=\{(X_i,M_i,y_i)\}^{N}_{i=1}$, be the training set of $N$ patients, where $X_i \in \mathbb{R}^{1xDxWxH}$ denotes a 3D NCCT scan with $D$ slices of height and width $H$ and $W$ respectively,  $M_i \in \mathbb{R}^{1xV}$ represents a metadata information vector of size $V$ with $V=2$ when $M_i$ is only treatment information and $V=52$ when $M_i$ is clinical metadata information, and finally, $y_i \in \{0,\dots,C-1\}$ corresponds to the $i^{th}$ label from $C$ classes, with  $C=2$ when the mRS score is dichotomised, and $C=7$ otherwise.

{\bf IFE module --} 
We need to extract local and global features from 3D volumes to make robust mRS predictions. To achieve this, we apply 3D convolution blocks to extract low and high level features, followed by squeeze \& excitation (SE) modules \cite{Hu2018Squeeze-and-ExcitationNetworks,Roy2018ConcurrentNetworks} that model global feature inter-dependencies both channel-wise and spatially (see cSE and sSE blocks in Fig. \ref{fig:mr-network-design}). Therefore, our network is encouraged to focus on important information through soft self-attention to learn more representative features at minimal computational cost.

The IFE module processes an input volume $X_i$ with three \textit{conv} blocks, each containing a 3D convolutional layer with a $3\times3\times3$ kernel size, followed by instance normalisation and LeakyRelu activation. The first two blocks also apply a downsampling layer with a strided-convolution.
We use instance normalisation over batch normalisation to be data frugal with memory requirements given our 3D {data}, and because a small batch size can lead to incorrect batch statistics which would decrease the model performance. Liu et al. \cite{liu2019design} have also recently reported on the benefits of instance normalisation over batch normalisation for 3D medical image processing.

After applying the channel-wise cSE and  spatial sSE attentional blocks to the high-level features from our final {\it conv} block, we add these recalibrated feature maps to the original high-level feature maps element-wise. Then, a global average pooling (GAP in Fig. \ref{fig:mr-network-design}) on the aggregated matrix, generates our final image features, $\Tilde{X}_i$, in preparation for fusion in the IMF module.

{\bf IMF module --} 
This module applies a fully connected layer to the image features $\Tilde{X}_i$ and the metadata information $M_{i}$, followed by a ReLU activation function, to generate encoded  image  volume  features, $F_{\Tilde{X}_i} \in \mathbb{R}^{1xJ}$ where $J$ is the feature vector size and metadata features, $F_{M_i}\in \mathbb{R}^{1xL}$ with $L\leq J$. These features are then concatenated to become the output of the IMF module, such that 
\begin{equation}
\begin{split}
    \Tilde{X}_i = \mbox{IFE}(X_i) ~, \\
    F_{\Tilde{X}_i,M_i}  = \mbox{IMF}(F_{\Tilde{X}_i}, F_{M_i}) ~.
\end{split}
\end{equation}

{\bf Generating the mRS score --} The classification stage of the proposed network processes the fused features through a FC layer with $C$ outputs. Finally, a Softmax function is applied to the outputs to generate a probability vector of $C$ classes,   
 $p_i = \{p^c_i\}^{C-1}_{c=0}$, where $p_i$ is the predicted mRS probability  scores for the $C$ classes, \ie 
\begin{equation}
    p_i = \mbox{softmax}( f (F_{\tilde{X}_i,M_i} )) ~, 
\end{equation} 
and the predicted mRS score is 
\begin{equation}
\mbox{mRS score } = \arg\max_{c} (p^c_i) . 
\end{equation}

{\bf Class Imbalance --}
Class imbalance leads to a bias toward the majority class in training data, and in the MR CLEAN Trial data, the distribution of the classes are highly imbalanced, \ie \{7, 36, 84, 87, 133, 45, 108\} patients for mRS scores 0 to 6, respectively. To deal with this data shortage, we use focal loss \cite{lin_focal_2017,abraham2019novel} to generate an adaptive weighting loss function, such that
\begin{equation}
    \mathcal{L_{\mathbf{FL}}}= -\sum_{i=1}^N \alpha_i y_i(1-p_i)^\gamma \log(p_i) ~,
    \label{eq:fl_loss}
\end{equation}
where $\alpha_i$ is a prefixed value between 0 and 1 to balance the positive and negative labelled samples. The focal intensity factor is $\gamma$ where if $\gamma =0$ and $\alpha_i =1$, the loss becomes a standard cross entropy loss.

Furthermore, as adopted in \cite{maier2017isles,HILBERT2019103516}, we train our networks and report our results both for all classes, and for groups of classes (\ie mRS scores $0-2$ as good outcome and mRS scores $3-6$ as undesirable outcome) which helps mitigate further for the class imbalance problem.

{\bf DNN for Clinical Metadata Only --} 
We also build a DNN for training clinical metadata information only to compare its predictive value with models that utilised image information. This DNN consists of two FC layers with hidden units of 128 and $C$ (2 or 7) number of output classes. We use dropout after the first FC layer for regularisation.

\section{Experiments and Results} \label{sec:exp_results}
{We purposed $80\%$ of our data (400 patients) for training and $20\%$ (100 patients) for testing. The full range of clinical metadata used in our experiments was the same as that in \cite{vanOs2018PredictingAlgorithms}.}

We evaluated the performance of our proposed network against other data and network configurations (in place of our IFE module). These are, (i) ResNet-18 and ResNet-34 as 3D implementations of ResNet \cite{he2016deep}, with their weights set by \textit{He normal} initialisation, (ii) ResNet-18$_{preT}$ and ResNet-34$_{preT}$ using pre-trained weights from \textit{MedicalNet} \cite{chen2019med3d} which were frozen (\textit{conv1 to layer4}) and then our IMF module was fine-tuned during training, (iii) U-NetEncoder, i.e. the encoder part of 3D U-Net \cite{cciccek20163d} initialised with \textit{He normal}, (iv) ClinicDNN, our own simple DNN applied on our clinical metadata only, and finally (v) the 3D CNN model from Bacchi \etal \cite{Bacchi2019DeepStudy}, trained from scratch, which generated mRS scores for thrombolysis (see Section \ref{sec:related_work} for details).

The networks in all our experiments were consigned with the same parameter values for training.  We set the $batch\_size$ to $8$ and number of $epochs$ to $300$ with early stopping if the network did not improve after 50 $epochs$. We used SGD with a momentum of $0.9$ as the optimizer. The learning rate was initialised to $0.00003$, determined empirically and decreased with cosine annealing scheduler. The experiments were performed on a single GPU (NVIDIA TITAN V 12GB) using PyTorch.

\begin{table}[t!]
\centering
        \caption{{Results for dichotomised mRS score ($0-2$ and $3-6$) when trained on 3D  volumes only, clinical metadata only and combination of 3D  volumes and metadata.}}
        \label{table:ths_mrs02FL}
\begin{tabular}{|l|l|c|c|c|c|c|c|}
\hline
                    & \textbf{Models}        & \textbf{ Accuracy }  &  {\it \textbf{F1-score} } & \textbf{ AUC }   \\ \hline
\multirow{6}{*}{\textbf{3D NCCT volume only}} & ResNet-18   & 0.74     & 0.00    & 0.58  \\
 &ResNet-18$_{preT}$ \cite{chen2019med3d}                  & 0.74     & 0.00    & 0.58    \\
&ResNet-34                                                  & 0.75     & 0.07    & 0.47   \\
&ResNet-34$_{preT}$ \cite{chen2019med3d}                    & 0.75     & 0.07    & 0.59   \\ 
&U-NetEncoder \cite{cciccek20163d}                           & 0.74     & 0.43    & 0.65   \\
&Bacchi \etal \cite{Bacchi2019DeepStudy}                    & 0.74     & 0.00    & 0.50    \\ 
&{Proposed Method}                                   & 0.75   & 0.46  & 0.67  \\
\hline
\hline
\textbf{Clinical metadata only} & Proposed ClinicDNN      & 0.66 & 0.56 & 0.70 \\
\hline
\hline
\multirow{6}{4cm}{\textbf{3D NCCT volume and Clinical metadata}} & ResNet-18   & 0.74            & 0.00          & 0.59             \\
 &ResNet-18$_{preT}$ \cite{chen2019med3d}                                      & 0.73            & 0.00          & 0.58             \\
&ResNet-34                                                                     & 0.74            & 0.00          & 0.52             \\
&ResNet-34$_{preT}$ \cite{chen2019med3d}                                       & 0.70            & 0.35          & 0.63             \\ 
&U-NetEncoder \cite{cciccek20163d}                           & 0.75     & 0.59    & 0.73   \\
&Bacchi \etal \cite{Bacchi2019DeepStudy}                                       & \textbf{0.78}   & 0.42          & 0.74             \\ 
&{Proposed Method}                                                      & 0.77           & \textbf{0.62}  & \textbf{0.75}    \\
\hline
\end{tabular}
\end{table}

We present two main experiments. In the first, {\it Exp.A}, we dichotomised the mRS scores into $C=2$ classes, good outcome ($\mbox{mRS} = 0-2$) and bad outcome ($\mbox{mRS} = 3-6$) of ischaemic stroke treatment. In {\it Exp.B}, we used all mRS scores individually for classification as $C=7$ classes ($\mbox{mRS} = 0,1,2,3,4,5,6$). We trained each network both on 3D NCCT volumes only and 3D NCCT volumes with clinical metadata, except for our simple ClinicDNN network which was trained only on clinical metadata. 

{\bf Exp. A - Dichotomous mRS scores --} 
The performance of the various models is compared by classification Accuracy, {\it F1-score} and Area Under ROC Curve (AUC) and their results are reported in Table \ref{table:ths_mrs02FL}. When training was based on 3D NCCT volumes only, our proposed model performed best, with $0.75$ Accuracy
, $0.46$ {\it F1-score} and $0.67$ AUC, while other networks performed significantly worse in {\it F1-score} and AUC. 

Our simple ClinicDNN network achieved 10\% better in {\it F1-score} and 3\% better in AUC than our model using 3D NCCT volumes only, but this can be attributed to the  radiological biomarkers, determined by expert radiologists from 3D NCCT scans, that were included in the clinical metadata. 

The leading results were obtained when combined 3D NCCT scans and clinical metadata were deployed for training. While Bacchi \etal \cite{Bacchi2019DeepStudy}'s model performed best in classification accuracy at $0.78$,  our proposed method produced the best results in {\it F1-score}, increasing from $0.46$ to $0.62$, and in AUC, climbing from $0.67$ to $0.75$, compared to when using 3D volume data only.

{\bfseries Exp. B - Individual mRS scores --} In this experiment, we predict each mRS score - 7 classes - individually. This means the expected accuracy would drop due to the scarcity of data in some classes, despite our class-imbalance mitigations, and since the groundtruth labels may not be precisely accurate given clinicians' judgements. Hence, we use both classification Accuracy and {1-Nearest Accuracy} as performance comparison measures where the latter is the percentage of samples either correctly classified or classified as the one nearest class (\ie correct if $y_i-1 \leq \arg\max(p_i) \leq y_i+1$). This allows a measure for a reasonable prediction of EVT outcome.

\begin{table}[t!]
\centering
        \caption{Results for individual mRS scores when models trained on 3D  volumes only, clinical metadata only and combination of 3D  volumes and clinical metadata.}
\label{table:mrs7_results}
\begin{tabular}{|l|l|c|c|}
\hline

        & \textbf{Models}       & \textbf{ Accuracy }   & \textbf{ \textit{1-Nearest Acc.}}  \\ \hline

\multirow{6}{*}{\textbf{3D NCCT volume only}} 
 & ResNet-18                                  &  0.30 & 0.63  \\
 &ResNet-18$_{preT}$ \cite{chen2019med3d}     & 0.25  & 0.57   \\
&ResNet-34                                    &  0.29 & 0.60   \\
&ResNet-34$_{preT}$ \cite{chen2019med3d}     & 0.31  & 0.61   \\
&U-NetEncoder \cite{cciccek20163d}            & 0.32  & 0.59   \\
&Bacchi \etal \cite{Bacchi2019DeepStudy}      & 0.27 & 0.56   \\ 
&Proposed method                     & 0.32 & \textbf{0.66}  \\
\hline
\hline
\textbf{Clinical metadata only} & Proposed ClinicDNN    & 0.29 & 0.64 \\
\hline
\hline
\multirow{6}{4cm}{\textbf{3D NCCT volume and Clinical metadata}} 
& ResNet-18                                  &  0.31  & 0.64  \\
 &ResNet-18$_{preT}$ \cite{chen2019med3d}   & 0.31  & 0.62   \\
&ResNet-34                                   &  0.29  & 0.63  \\
&ResNet-34$_{preT}$ \cite{chen2019med3d}     & 0.32  & 0.55  \\
&U-NetEncoder \cite{cciccek20163d}           & 0.32   & 0.60   \\
&Bacchi \etal \cite{Bacchi2019DeepStudy}     & 0.24 & 0.57   \\ 

& Proposed method                    & \textbf{0.35} & 0.63 \\
\hline
\end{tabular}
\end{table}

The results of this experiment are reported in Table \ref{table:mrs7_results}. Our proposed method performed best in both comparison metrics compared to other methods, albeit when using different data types.
Similar to {\it Exp.A}, the performance of many of the models benefited from combining imaging and clinical metadata information. {However, our proposed model's performance decreased in the {1-Nearest Accuracy} measure when using multimodal data, since for individual mRS score classification, the predictions were spread across classes instead of accumulating around the majority ones (mRS 4 and 6).} 

{\bf Ablation Study -- } Table \ref{table:ablation} demonstrates the importance of the attention modules, cSE and sSE, when they are omitted from our proposed method. The performance of our model decreased for example, in classification accuracy from $0.77$ down to $0.71$, in {\it F1-score} from $0.62$ down to $0.57$ and in AUC from $0.75$ down to $0.72$, in the case of dichotomised mRS scores and combination of imaging and clinical metadata information.

\begin{table}[t!]
\centering
        \caption{Ablation study results when attention modules are not deployed.}
\label{table:ablation}
\begin{tabular}{|l|c|c|c|c|c|}
\hline
\multirow{2}{*}{} & \multicolumn{3}{c|}{\textbf{Dichotomised mRS}}   & \multicolumn{2}{c|}{\textbf{Individual mRS}}  \\ \cline{2-6}
       &Accuracy &{\it F1-score} &  AUC  &Accuracy&{\it 1-Nearest Acc.} \\ \hline
\textbf{3D NCCT volume only }   & 0.67  & 0.41 &  0.62 &  0.27 & 0.63   \\
\hline
\begin{tabular}[c]{@{}l@{}}\textbf{3D NCCT volume and} \\ \textbf{Clinical metadata}\end{tabular} & 0.71 & 0.57 & 0.72& 0.29& 0.60 \\
\hline
\end{tabular}
\end{table}

\section{Conclusions}  \label{sec:conclusions}
In this study, we presented a novel approach to predicting functional outcome of ischaemic stroke treatment from multimodal data. We added an attention mechanism in our architecture to extract representative features both spatially and channel-wise. In both {\it Exp.A} and {\it Exp.B}, our proposed method outperformed other methods,  with $0.62$ {\it F1-score} and $0.75$ AUC and in dichotomous  mRS scores and $0.35$ classification Accuracy in individual mRS scores, leading us to conclude that  combining clinical metadata information with 3D imaging information and attention modules increases performance. 

The results also demonstrate that a well-designed shallower network can perform better than deep networks for this specific task, given the architectures of our proposed model and Bacchi \etal (in  Exp.A) against the ResNet models, while the deep ResNet models perform better than Bacchi et al. in Exp.B, potentially due to their better generalisation in the face of class imbalance. 

For future work, we plan to investigate the class imbalance issue more elaborately, while collecting more data for the minority classes. We aim to explore other data fusion strategies, \eg adaptive fusion and late fusion. Given sufficient data, we plan to also analyse temporal changes in stroke regions in 3D NCCT  follow-up scans for better prediction of functional outcome of treatment. 

\section*{Acknowledgements} 
{The authors would like to thank the MR CLEAN Registry team: Prof Aad van der Lugt, Prof Diederik W.J. Dippel, Prof. Charles B.L.M. Majoie, Prof. Wim H. van Zwam and Prof. Robert J. van Oostenbrugge for providing the data.
Zeynel Samak gratefully acknowledges funding from Ministry of Education (1416/YLSY), the Republic of Turkey. The Titan V used for this research was donated by the NVIDIA Corporation.}

\bibliographystyle{splncs04}
\bibliography{short_references}

\begin{thebibliography}{10}
\providecommand{\url}[1]{\texttt{#1}}
\providecommand{\urlprefix}{URL }
\providecommand{\doi}[1]{https://doi.org/#1}

\bibitem{abraham2019novel}
Abraham, N., Khan, N.M.: {A novel focal tversky loss function with improved
  attention U-Net for lesion segmentation}. In: ISBI. pp. 683--687. IEEE (2019)

\bibitem{albers2018thrombectomy}
Albers, G.W., \etal: Thrombectomy for stroke at 6 to 16 hours with selection by
  perfusion imaging. NEJM  \textbf{378}(8),  708--718 (2018)

\bibitem{Asadi2014}
Asadi, H., \etal: {Machine Learning for Outcome Prediction of Acute Ischemic
  Stroke Post Intra-Arterial Therapy}. PLoS ONE  \textbf{9}(2),  e88225 (2014)

\bibitem{Bacchi2019DeepStudy}
Bacchi, S., \etal: {Deep Learning in the Prediction of Ischaemic Stroke
  Thrombolysis Functional Outcomes: A Pilot Study}. Academic Radiology  (2019)

\bibitem{Bentley2014}
Bentley, P., \etal: {Prediction of stroke thrombolysis outcome using CT brain
  machine learning}. NeuroImage: Clinical  \textbf{4},  635--640 (2014)

\bibitem{Berkhemer2015AStroke}
Berkhemer, O.A., \etal: {A randomized trial of intraarterial treatment for
  acute ischemic stroke}. NEJM  \textbf{372}(1),  11--20 (2015)

\bibitem{Boers2013AutomatedStroke}
Boers, A., \etal: {Automated Cerebral Infarct Volume Measurement in Follow-up
  Noncontrast CT Scans of Patients with Acute Ischemic Stroke}. AJN
  \textbf{34}(8),  1522--1527 (2013)

\bibitem{bohme2018combining}
B{\"o}hme, L., \etal: {Combining Good Old Random Forest and DeepLabv3+ for
  ISLES 2018 CT-Based Stroke Segmentation}. In: MICCAIBW. pp. 335--342.
  Springer (2018)

\bibitem{Chawla2009AImages}
Chawla, M., \etal: {A method for automatic detection and classification of
  stroke from brain CT images}. In: IEEEMBS. pp. 3581--3584. IEEE (2009)

\bibitem{chen2019med3d}
Chen, S., \etal: {Med3D: Transfer Learning for 3D Medical Image Analysis}.
  arXiv preprint arXiv:1904.00625  (2019)

\bibitem{choi2016ensemble}
Choi, Y., \etal: {Ensemble of deep convolutional neural networks for prognosis
  of ischemic stroke}. In: IWBGMSSTBI. pp. 231--243. Springer (2016)

\bibitem{cciccek20163d}
{\c{C}}i{\c{c}}ek, {\"O}., \etal: {3D U-Net: learning dense volumetric
  segmentation from sparse annotation}. In: MICCAI. pp. 424--432. Springer
  (2016)

\bibitem{fahed2018dwi}
Fahed, R., \etal: Dwi-aspects (diffusion-weighted imaging--alberta stroke
  program early computed tomography scores) and dwi-flair (diffusion-weighted
  imaging--fluid attenuated inversion recovery) mismatch in thrombectomy
  candidates: An intrarater and interrater agreement study. Stroke
  \textbf{49}(1),  223--227 (2018)

\bibitem{forkert2015multiclass}
Forkert, N.D., \etal: {Multiclass support vector machine-based lesion mapping
  predicts functional outcome in ischemic stroke patients}. PLoS One
  \textbf{10}(6) (2015)

\bibitem{fransen2014mr}
Fransen, P.S., \etal: {MR CLEAN, a multicenter randomized clinical trial of
  endovascular treatment for acute ischemic stroke in the Netherlands: study
  protocol for a randomized controlled trial}. Trials  \textbf{15}(1), ~343
  (2014)

\bibitem{GOYAL20161723}
Goyal, M., \etal: {Endovascular thrombectomy after large-vessel ischaemic
  stroke: a meta-analysis of individual patient data from five randomised
  trials}. The Lancet  \textbf{387}(10029),  1723--1731 (2016)

\bibitem{gupta2014brain}
Gupta, N., Mittal, A.: {Brain ischemic stroke segmentation: a survey}. Journal
  of Multi Disciplinary Engineering Technologies Volume  \textbf{8}(1), ~1
  (2014)

\bibitem{he2016deep}
He, K., \etal: {Deep residual learning for image recognition}. In: CVPR (2016)

\bibitem{heo2018machine}
Heo, J., \etal: {Machine learning-based model can predict stroke outcome}.
  Stroke  \textbf{49}(Suppl\_1),  A194--A194 (2018)

\bibitem{HILBERT2019103516}
Hilbert, A., \etal: {Data-efficient deep learning of radiological image data
  for outcome prediction after endovascular treatment of patients with acute
  ischemic stroke}. Computers in Biology and Medicine p. 103516 (2019)

\bibitem{Hu2018Squeeze-and-ExcitationNetworks}
Hu, J., \etal: {Squeeze-and-Excitation Networks}. In: CVPR (2018)

\bibitem{isensee2017brain}
Isensee, F., \etal: {Brain tumor segmentation and radiomics survival
  prediction: Contribution to BRATS 2017 challenge}. In: MICCAIBW. pp. 287--297
  (2017)

\bibitem{jansen2018endovascular}
Jansen, I.G., \etal: {Endovascular treatment for acute ischaemic stroke in
  routine clinical practice: prospective, observational cohort study (MR CLEAN
  Registry)}. BMJ  \textbf{360}, ~k949 (2018)

\bibitem{kamnitsas2017ensembles}
Kamnitsas, K., \etal: {Ensembles of multiple models and architectures for
  robust brain tumour segmentation}. In: MICCAIBW. pp. 450--462. Springer
  (2017)

\bibitem{lin_focal_2017}
Lin, T.Y., \etal: {Focal loss for dense object detection}. In: CVPR (2017)

\bibitem{lisowska2017context}
Lisowska, A., \etal: {Context-aware convolutional neural networks for stroke
  sign detection in non-contrast CT scans}. In: MIUA. pp. 494--505. Springer
  (2017)

\bibitem{liu2019design}
Liu, S., \etal: {On the design of convolutional neural networks for automatic
  detection of Alzheimer's disease}. In: NeurIPS ML4H (2019)

\bibitem{Maier2014IschemicClassifiers}
Maier, O., \etal: {Ischemic stroke lesion segmentation in multi-spectral MR
  images with support vector machine classifiers}. In: Medical Imaging 2014:
  Computer-Aided Diagnosis. vol.~9035, p. 903504. ISOP (2014)

\bibitem{maier2017isles}
Maier, O., \etal: {ISLES 2015-A public evaluation benchmark for ischemic stroke
  lesion segmentation from multispectral MRI}. MIA  \textbf{35},  250--269
  (2017)

\bibitem{Maier2016PredictingForests}
Maier, O., Handels, H.: {Predicting Stroke Lesion and Clinical Outcome with
  Random Forests}. In: MICCAI Brainlesion Workshop. pp. 219--230. Springer
  (2016)

\bibitem{matesin2001rule}
Matesin, M., \etal: {A rule-based approach to stroke lesion analysis from CT
  brain images}. In: ISPA. pp. 219--223. IEEE (2001)

\bibitem{mckinley2017fully}
McKinley, R., \etal: {Fully automated stroke tissue estimation using random
  forest classifiers (FASTER)}. JCBFM  \textbf{37}(8),  2728--2741 (2017)

\bibitem{nishi2019predicting}
Nishi, H., \etal: {Predicting Clinical Outcomes of Large Vessel Occlusion
  Before Mechanical Thrombectomy Using Machine Learning}. Stroke
  \textbf{50}(9),  2379--2388 (2019)

\bibitem{parisot2018disease}
Parisot, S., \etal: {Disease prediction using graph convolutional networks:
  Application to Autism Spectrum Disorder and Alzheimer’s disease}. MIA
  \textbf{48},  117--130 (2018)

\bibitem{pinto2018enhancing}
Pinto, A., \etal: {Enhancing clinical MRI Perfusion maps with data-driven maps
  of complementary nature for lesion outcome prediction}. In: MICCAI. pp.
  107--115. Springer (2018)

\bibitem{Rekik2012MedicalAppraisal}
Rekik, I., \etal: {Medical image analysis methods in MR/CT-imaged
  acute-subacute ischemic stroke lesion: Segmentation, prediction and insights
  into dynamic evolution simulation models. A critical appraisal}. NeuroImage:
  Clinical  \textbf{1}(1),  164--178 (2012)

\bibitem{renowden2014imaging}
Renowden, S.: {Imaging in stroke and vascular disease—part 1: ischaemic
  stroke}. Practical Neurology  \textbf{14}(2),  77--87 (2014)

\bibitem{Roy2018ConcurrentNetworks}
Roy, A.G., \etal: {Concurrent Spatial and Channel ‘Squeeze {\&}amp;
  Excitation’ in Fully Convolutional Networks}. In: MICCAI. pp. 421--429.
  Springer (2018)

\bibitem{StrokeAssociation2018StateStatistics}
{Stroke Association}: {State of the Nation: stroke statistics} (2018),
  \url{https://www.stroke.org.uk/resources/state-nation-stroke-statistics}
  [Accessed Nov-2019]

\bibitem{vanOs2018PredictingAlgorithms}
Van~Os, H.J., \etal: {Predicting outcome of endovascular treatment for acute
  ischemic stroke: potential value of machine learning algorithms}. Frontiers
  in Neurology  \textbf{9}, ~784 (2018)

\bibitem{Venemaj1710}
Venema, E., \etal: Selection of patients for intra-arterial treatment for acute
  ischaemic stroke: development and validation of a clinical decision tool in
  two randomised trials. BMJ  \textbf{357} (2017)

\bibitem{weimar2002predicting}
Weimar, C., \etal: Predicting functional outcome and survival after acute
  ischemic stroke. Journal of Neurology  \textbf{249}(7),  888--895 (2002)

\bibitem{WHO2018TheDeath}
{WHO}: {The top 10 causes of death} (2018),
  \url{https://www.who.int/en/news-room/fact-sheets/detail/the-top-10-causes-of-death}
  [Accessed Nov-2019]

\bibitem{Winzeck2018ISLESMRI.}
Winzeck, S., \etal: {ISLES 2016 and 2017-Benchmarking Ischemic Stroke Lesion
  Outcome Prediction Based on Multispectral MRI}. Frontiers in Neurology
  \textbf{9} (2018)

\end{thebibliography}

\end{document}